# High-speed programmable photonic circuits in a cryogenically compatible, visible-NIR 200 mm CMOS architecture

Mark Dong,[1,2,6] Genevieve Clark,[1,2] Andrew J. Leenheer,[3] Matthew Zimmermann,[1] Daniel Dominguez,[3] Adrian J. Menssen,[2] David Heim,[1] Gerald Gilbert,[2,4,7] Dirk Englund,[2,5,8] and Matt Eichenfield[3,9]

[1]*The MITRE Corporation, 202 Burlington Road, Bedford, Massachusetts 01730, USA*
[2]*Research Laboratory of Electronics, Massachusetts Institute of Technology, Cambridge, Massachusetts 02139, USA*
[3]*Sandia National Laboratories, P.O. Box 5800 Albuquerque, New Mexico, 87185, USA*
[4]*The MITRE Corporation, 200 Forrestal Road, Princeton, New Jersey 08540, USA*
[5]*Brookhaven National Laboratory, 98 Rochester St, Upton, New York 11973, USA*
[6]*mdong@mitre.org*
[7]*ggilbert@mitre.org*
[8]*englund@mit.edu*
[9]*meichen@sandia.gov*

## Summary

Recent advances in photonic integrated circuits (PICs) have enabled a new generation of "programmable many-mode interferometers" (PMMIs) realized by cascaded Mach Zehnder Interferometers (MZIs) capable of universal linear-optical transformations on *N* input-output optical modes. PMMIs serve critical functions in photonic quantum information processing, quantum-enhanced sensor networks, machine learning and other applications. However, PMMI implementations reported to date rely on thermo-optic phase shifters, which limit applications due to slow response times and high power consumption. Here, we introduce a large-scale PMMI platform, based on a 200 mm CMOS process, that uses aluminum nitride (AlN) piezo-optomechanical actuators coupled to silicon nitride (SiN) waveguides, enabling low-loss propagation with phase modulation at greater than 100 MHz in the visible to near-infrared wavelengths. Moreover, the vanishingly low holding-power consumption of the piezo-actuators enables these PICs to operate at cryogenic temperatures, paving the way for a fully integrated device architecture for a range of quantum applications.

## Introduction

For many applications proposed for programmable optical systems,[1–3] the key requirements are (i) large scale , (ii) fast response -- sub-µs for applications in machine learning,[4–7] quantum control,[8–11] and (iii) power efficiency to enable operation at cryogenic environment for integration with superconducting detectors[12,13] and artificial atoms.[14–16] Leading PMMI platforms[17] consist of cascaded MZIs to perform the general SU($N$) unitary requiring $N(N-1)/2$ MZIs,[18] as illustrated in Fig.1a for a 4-mode programmable SU(4) in a rectangular mesh arrangement.[19] To meet the scalability requirement (i), for even modest mode count $N$, the most demanding technical challenge is reliable manufacturing and a clear path to electronics co-integration. As presently only modern CMOS-compatible VLSI processes offer these capabilities through foundry services, material choices are practically limited to Si[20,21] and silicon nitride technologies.[22] For the high-bandwidth requirement (ii) within these materials, silicon free carrier modulation[23] is problematic for PMMIs because of inherently coupled phase and amplitude. Silicon electric-field induced $\chi^{(2)}$[24] modulators are promising but have not been realized at scale and restrict operating wavelength power. A recently introduced alternative relies on piezo-optomechanical actuation of SiN waveguides,[25–28] which in the case of [25] has enabled narrow-band operation up to 250 MHz with SiN waveguides for visible-NIR operation and high

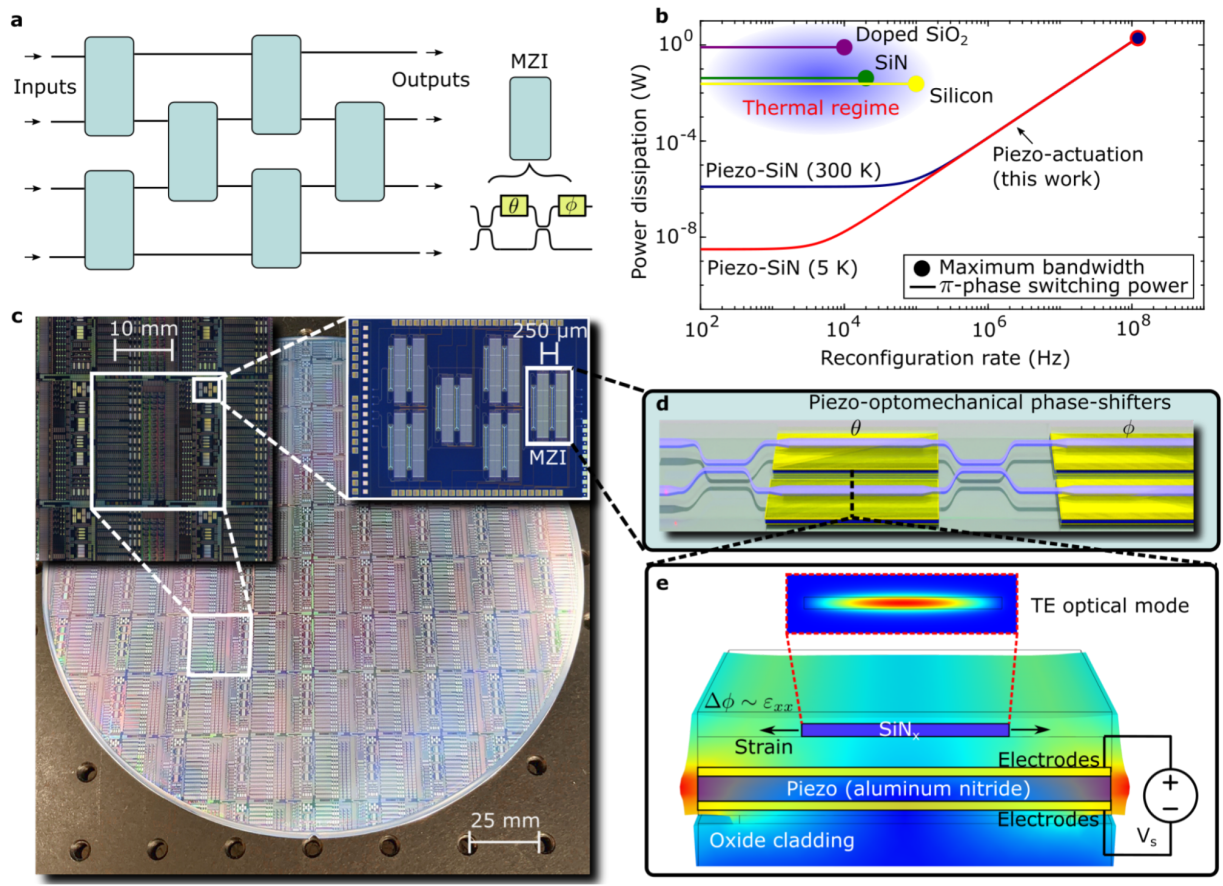

**Fig. 1: High-speed, visible-wavelength and cryogenic-compatible photonic integrated circuits.** a) schematic of 4-mode programmable circuit consisting of cascaded MZIs for SU(4) operations b) power-reconfiguration rate trade space diagram for thermally-actuated CMOS-compatible photonic technologies and our piezo-actuated photonics c) microscope images of a fully processed 200 mm wafer with zoomed insets showing a reticle and a 4x4 PMMI d) 3D concept art of an MZI with piezo-optomechanical phase-shifters e) cross-sectional diagram of our phase-shifter, illustrating how an applied voltage across the piezo results in strain imparted to the optical waveguide

optical power handling. Finally, the energy requirement (iii) is incompatible with any thermal modulation schemes, which dissipate an average of >800 mW for doped SiO2,[21] >20 mW per modulator for SOI[29], and >40 mW per modulator for SiN.[17] Electrostatic MEMS devices meet the power requirement,[30,31] but modulation timescales have thus far been limited to the µs regime.

Fig. 1 (b) summarizes this power-modulation bandwidth trade-space for SiN and Si platforms compatible with requirement (i). For cross-platform comparison, the figure plots the power dissipated for π-phase shifts as a function of circuit reconfiguration rate. In the thermal regime, the power dissipated is constant up to the maximum reconfiguration bandwidth, while piezo-actuation contains a trade-off between faster programming speeds and power dissipated on-chip. The piezo-optomechanical approach offers up to three orders of magnitude improvement in programming time with less power consumption as that of thermal and three to five orders of magnitude improvement in power dissipation at the maximum thermal reconfigurability rate, showing the exceptional suitability of this technology for PMMI systems. Motivated by these considerations, we advance our wafer-scale process for SiN PICs (Fig. 1c) with phase-only piezo-optomechanical tuners, whose basic design and operation are shown in Fig. 1d, e, through co-design and development of photonics, electromechanics, electrical and mechanical control systems, and driver software. We combine the best attributes of different devices in our previous work[25] into a proof-of-principle PMMI architecture capable of 20 ns reconfiguration time, on-chip dissipation per modulator below 200 µW when switching on average every 1 µs and 6 nW power to hold, operation at 700 - 780 nm with optical transmission up to 1550 nm, and improved power efficiency at 5 K.

## 2.   Results

### A. Device fabrication, design, and theory of operation

Our devices are designed and fabricated on a 200 mm wafer following an improved process flow based on our previous work.[25] The optical layer consists of a 300 nm thick silicon nitride (SiN) layer in a silicon dioxide cladding, located atop the AlN piezo-electric actuators with functionality for selective release and metal routing; see Supplementary Section 1 for more fabrication details. This process enables high-speed and broadband active modulators. Specifically, Fig.2a shows a full MZI with two strain-actuated phase-shifters internal and external to the directional couplers, as in the unfolded layout of Fig.1d. Each phase-shifter consists of a 400 nm width SiN waveguide passing through an adiabatic taper, expanding to 5 µm width for improved strain transfer,[25] and then through a second adiabatic taper back down to 400 nm-wide single-mode turn. This loop repeats nine times as shown in Fig. 2b. The cross-section in Fig. 2c, d shows a slightly undercut 10 µm-wide pillar around the oxide-clad 5 µm-wide SiN waveguide and AlN actuators. Applying a potential difference across the AlN transfers strain to the waveguide (see finite-element simulation in Fig. 2e), imparting a refractive index shift predominantly due to the strain-optic effect[32] in addition to moving boundary effects.[33]

When applying a potential difference, the strain imparts a phase change $\Delta\theta$ to the optical mode for a given length $L$ phase-shifter is

$$\Delta\theta = (2\pi/\lambda)(\Delta n_{eff} L + n_{eff} \Delta L), \quad (1)$$

where $\lambda$ is the free-space wavelength, $n_{eff}$ is the static effective mode index, $\Delta L$ is the strain-induced path-length change of the phase-shifter, and $\Delta n_{eff}$ is the total effective strain-induced refractive index change:

$$\Delta n_{eff} = \Delta n_{eff,\varepsilon} + \Delta n_{eff,wg} \quad (2)$$

$$\Delta n_{eff,\varepsilon} = -p_{11}(n^3_{eff}/2)\varepsilon_{xx} \quad (3)$$

which includes contributions from the strain-optic and the moving boundary effects. Here, $p_{11}$ is the diagonal component of the strain-optic coefficient for SiN, and $\varepsilon_{xx}$ is the tensile strain component in the horizontal ($x$) direction parallel to the substrate surface (labeled in Fig. 2e). We estimate the strain values transferred to the SiN from a finite-element model using approximate values of AlN piezoelectric coupling coefficients,[25] layer stack geometries, and material properties, indicating that the dominant strain is tensile in the horizontal (*x*) direction and of magnitude $\varepsilon_{xx} \sim 1.5 \times 10^{-6}$ per volt applied. The change in effective index due to strain in Equation 3 is a signed quantity depending on the direction of the induced strain, meaning opposite strains produced by a negative applied voltage produce an opposite-shift effective index. This effect allows for a push-pull or differential operation of the phase-shifts $\Theta_{1,2}$, $\Phi_{1,2} \in [-\pi/2, \pi/2]$ in Fig. 2a for an applied $V_s \in [-V_\pi/2, V_\pi/2]$, defined as the potential relative to ground.

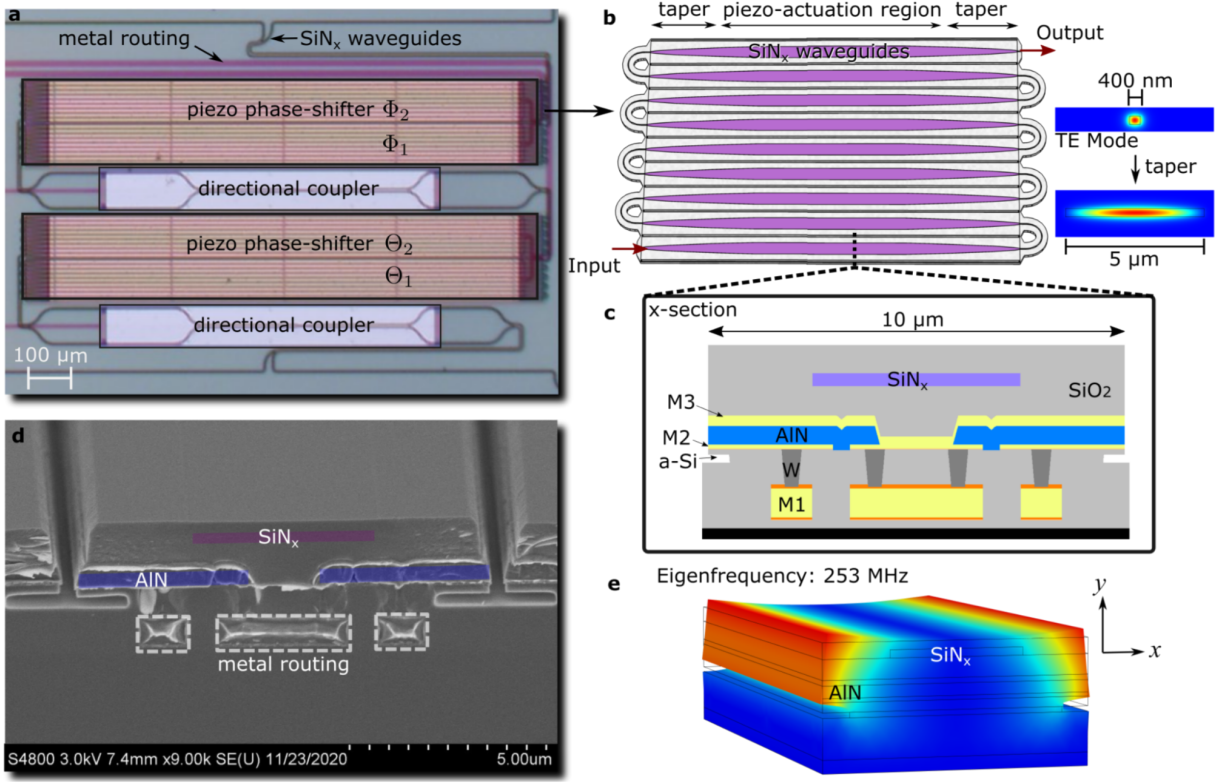

**Fig. 2: Design of piezo-actuated modulators** a) microscope photo of an MZI with labeled silicon nitride waveguides, metal routing (M1), phase-shifters and directional couplers b) model of phase-shifter with labeled optical input / output, waveguide taper and actuation regions with finite element simulations of the 400 nm and 5 μm wide waveguide TE modes c) schematic cross-section of phase-shifter with materials labeled d) SEM cross-section of fabricated device, with waveguides (purple), AlN (blue), and metal routing (white) labeled e) finite element simulation of the first mechanical eigenfrequency at 253 MHz with plotted displacement exaggerated for visibility.

### B. Room temperature operation

Fig. 3 summarizes the modulator performance at room temperature (300 K). Fig. 3a, b show operation across a broadband range of wavelengths (700 nm- 780 nm). The plots shown here take advantage of the push-pull operation for optical loss balance (-3.5 dB typical insertion loss at 737 nm, see Supplementary Section 2 for loss characterization details) of the phase-shifters by applying opposite polarity voltages to each arm. We estimate the $V_\pi L$ from sinusoidal fits, yielding values from 50 $V \cdot$cm (700 nm) to 65 $V \cdot$cm (780 nm) in the push-pull

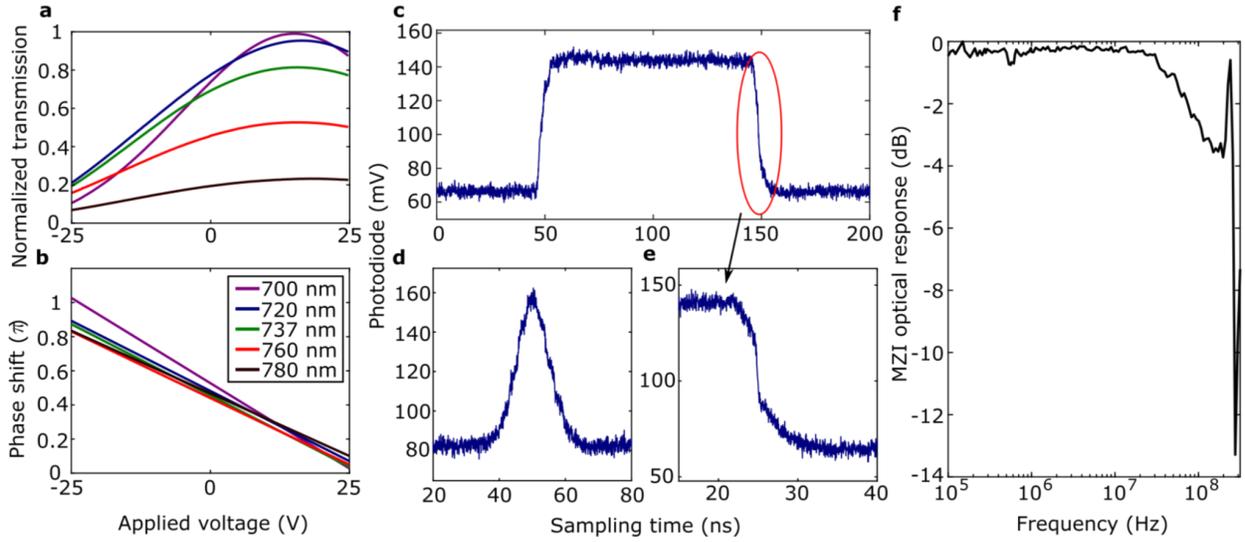

**Fig. 3: MZI modulator characterization at 300 K.** a) bar port power transmission coefficient of several wavelengths vs differential driving voltage of the MZI b) extracted phase-shifts at the same wavelengths c) photodiode response at 737 nm with a 5 MHz, 20 V pk-pk square wave applied d) photodiode output with ~20 ns, 20 V pk-pk gaussian pulse applied e) falling edge of 5 MHz square is zoomed in showing a ~5 ns fall time f) normalized modulator response at 3V sine signal showing a mechanical resonance at 241 MHz

configuration or a corresponding 100 $V \cdot cm$ to 130 $V \cdot cm$ $V_\pi L$ per single phase-shifter. We attribute the variance in $V_\pi L$ to material and waveguide dispersion.

Time-resolved measurements of Fig. 3c, d, e indicate short rise and fall times of ~5 ns, respectively. Fig. 3c plots the transmission when averaging over 16 square wave voltage pulses applied through an arbitrary waveform generator (AWG), indicating a fall time of 5 ns. As seen in Fig.3d, the modulator follows a Gaussian pulse of 15 ns FWHM. Fig. 3f plots the small-signal frequency-resolved modulator response, indicating a -3dB cutoff at $\nu_{3dB} = 120$ MHz. This cutoff is consistent with the RC constant during measurement, consisting of the device capacitance of 17 pF and series resistances of ~80 Ω arising from the voltage source and on-chip routing metal. We attribute the peak at 241 MHz to the fundamental mechanical resonance, which our finite-element model puts at 253 MHz (Fig. 2e). The shape of the peak exhibits a characteristic resonance and anti-resonance feature typical of that in mechanical resonators.[34,35] We note that the AWG used for these tests does not produce the full 20 V swing across its entire bandwidth (500 MHz) which accounts for the atypical falling edge and distorted Gaussian envelope seen in Fig. 3e, d respectively. The contrast in these measurements is low, as this device is optimized for cryogenic operation, as described next.

## C. Cryogenic operation at 5 K

We performed the cryogenic characterization in a 4 K closed-cycle cryostat (Montana Instruments). Using a fiber array and RF probe mounted on nanopositioners inside the cryostat (see Supplementary section 5), we applied various voltage waveforms. The transmission curves in Fig. 4 show no degradation in switching performance at a base temperature of 5 K (Fig. 4a, b) compared to room temperature. Fig. 4c indicates an extinction of approximately 30 dB. Fig. 4d plots a time trace when holding $V_s = 20$V, indicating very stable open-loop operation (<0.5% drift) on the minutes time-scale. The corresponding holding-power of 2 nW is nearly 8 orders of magnitude smaller than the Montana cryostat cooling power (90 mW). This low power consumption is a critical benefit, as even state of the art thermal SiN WG modulators (Fig. 1b) would easily overwhelm the cryostat's cooling power. We note the actual extinction ratio of the modulator is larger than 30 dB; however, limited volume in the cryostat prevented the use of differential driving as demonstrated at room temperature and limited the phase change to approximately $\pi/2$, rather than $\pi$.

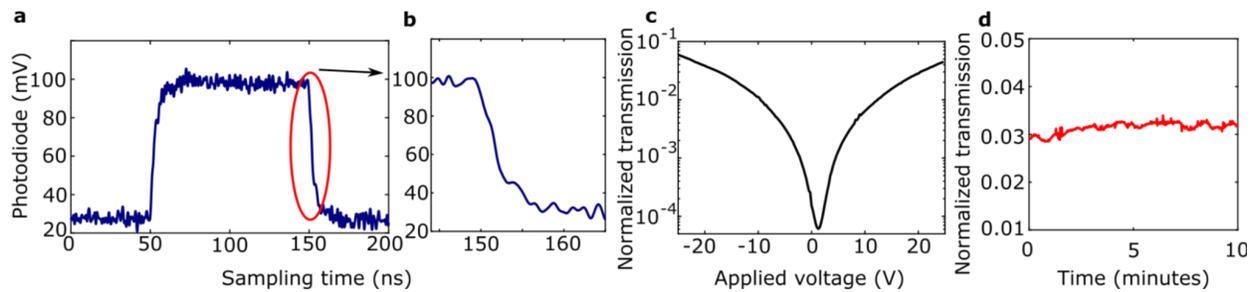

**Fig. 4: Cryogenic operation of the modulator at 5 K at 737 nm wavelength.** a) 16-sample averaged photodiode response for a 5 MHz, 20 V pk-pk square wave driving signal b) zoomed-in falling edge of square wave, showing the same ~5 ns fall time as that at room temperature c) DC sweep of a single phase-shifter in our modulator, showing a maximum extinction ratio of ~30 dB d) stability of our modulator held at 5 K with a constant 20 V applied, showing <0.5% drift of the MZI set-point over several minutes.

### D. 4x4 programmable interferometer

Having characterized the MZIs, we now consider a proof-of-principle SU(4) programmable unitary transformation, which consists of $N(N-1)/2 = 6$ MZIs. The PMMI is fully programmable, with all phase-shifters computer-controlled (see Supplementary Section 3).

Fig. 5 summarizes the PMMI characterization. After a basic voltage calibration, we run voltage sweeps across all six MZIs, producing the internal phase-shifter transmissions shown in Fig. 5d. The optical input and output combinations (labeled in Fig. 5c) here for each plotted MZI show maximum power differences when actuating – a full data set of all twelve phase-shifters is shown in Supplementary Section 4. Fig. 5d plots the transmission of an exemplary MZI: "MZI3," which indicates an optical extinction ratio in excess of 40 dB. The total loss through the PMMI ranges from -14 dB to -21 dB at 737 nm, depending on the optical path taken, primarily due to waveguide bending losses in the phase-shifters, which could be completely eliminated with an unfolded implementation or larger radii waveguide bends. We confirm that our PMMI maintains the 100 MHz bandwidth as demonstrated in the single MZI devices by applying a 10 V, 100 MHz sinusoid to the internal phase-shifter of MZI2 and monitor the optical outputs 5, 6 with a laser coupled through optical input 3. We utilize a lock-in detection scheme on the output photodiodes to extract the signal at 100 MHz (Fig. 5e, f) for both channels and a relative phase offset of $0.59\pi$; the small deviation from $\pi/2$ is likely due to the 125 MHz bandwidth of the photodiode being close to the driving frequency. Although results in Fig. 5 demonstrate the programmability of our PMMI as a single packaged device, voltage handling limitations of our on-chip vias (see Discussion section below) currently prevent accessing the full scope of U(*4*) operations.

### Discussion

We have demonstrated a 4-mode programmable interferometer in a 200 mm wafer-scale CMOS process. The PMMI is comprised of cryogenically compatible, >100 MHz bandwidth piezo-optomechanical phase-shifters operating in the visible to NIR wavelengths, which should be immediately applicable to many hybrid photonic architectures,[16] quantum protocols,[8,36] and machine-learning algorithms.[6]

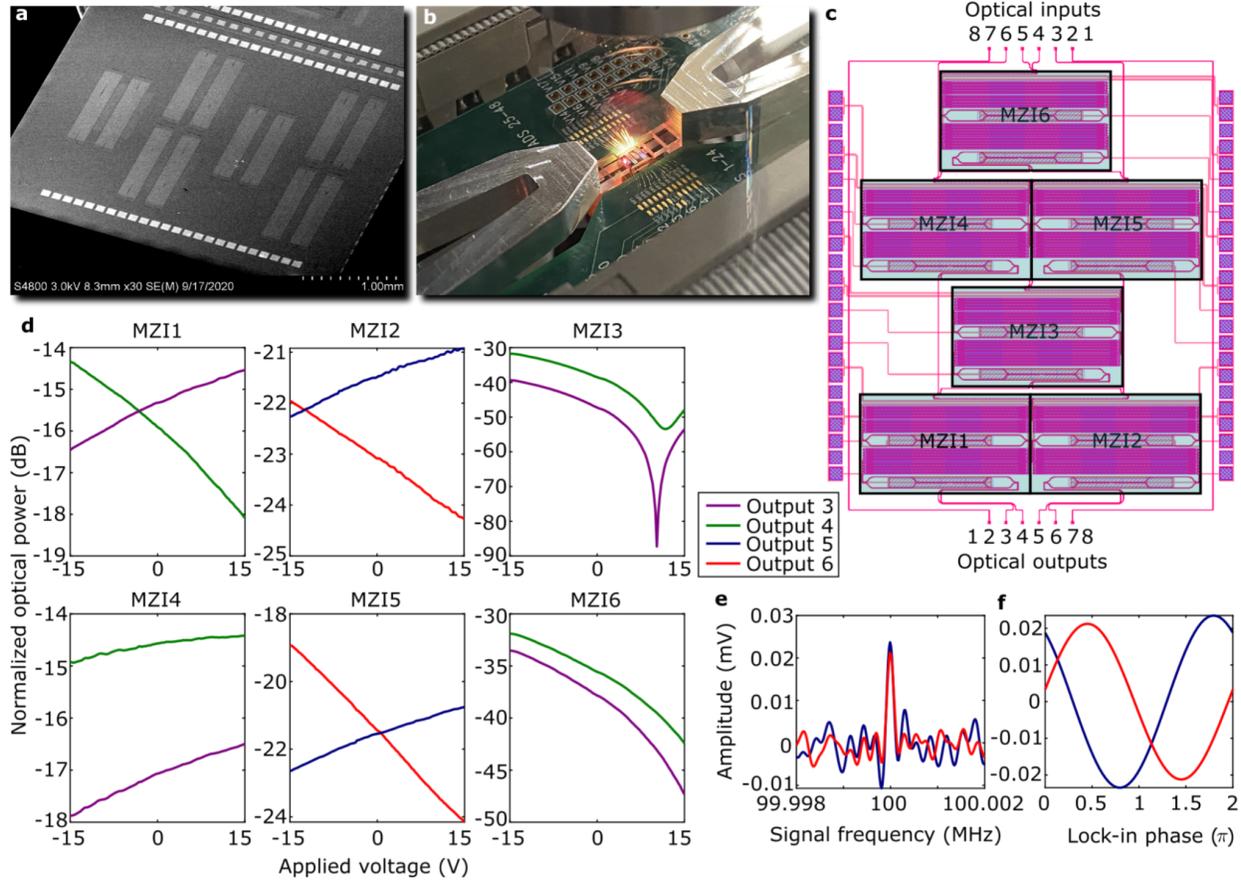

**Fig. 5: Proof of principle PMMI operation at 737 nm** a) SEM image of the same device b) image of photonic chip, wire bonds, PCB, and optical I/O for testing c) diagram of MZI list and labels for optical I/O; fiber channels 1, 8 are not connected and channels 2,7 are straight-through waveguides for alignment and calibration d) actuation of numbered MZI while plotting optical outputs 3,4 or 5,6; optical inputs used are input 3 for MZIs 2, 3, 5, input 4 for MZI6, and input 6 for MZI1; Plotted optical power is normalized to input laser power after accounting for facet coupling losses e) measured signal amplitude from fiber output 5, 6 of the 4x4 PMMI with a 10 V, 100 MHz driving sinusoid on MZI  b) signal amplitude as a function of lock-in phase; integration time for lock-in measurements is 4 ms

While the devices presented here show promise for universal linear-optic programmable photonics in new performance regimes and application spaces, there are still some non-idealities that will be improved in future work. First and foremost, the dominant source of loss in all devices presented is due to bends in waveguide meanders and directional couplers, measured to be -2.7 dB of the -3.5 dB MZI insertion loss (see Supplementary Section 2). Waveguide propagation losses (-0.39 dB/cm) and adiabatic taper losses -0.022 dB ± 0.038 dB per taper, account for the remaining -0.8 dB of loss. We estimate that implementing less tightly folded versions of the same devices that have fewer bends and tapers would allow us to achieve MZI insertion losses significantly less than -1 dB. Second, in these particular devices, small voids in the AlN in the vicinity of electrical vias produced metal filamentation that degraded the device breakdown voltage and prevented application of voltages above $V_s$ =25 V. While we found individual vias in test structures on the same wafer that allowed $V_s$ >90 V, the yield was insufficient for an MZI comprising 90 vias . After systematic failure analysis and cross-sectional imaging, the process yield improved greatly so that full SU(*N*) devices appear likely with $V_s$ >90 V. Moreover, high-scandium-concentration $Sc_{(1-x)}Al_xN$ has shown a five-fold increase in the piezoelectric responsivity,[37] promising a further 5 × reduction in $V_\pi$.

Finally, the layer-stack of our PIC (Table S1 and Fig. 2c), which places the optical layers on top, enables many extensions to the platform. The optical layer, being above all metal and piezoelectric layers, allows for alternative optical layers to expand the transparency window into the UV regime.[38–40] Moreover, heterogeneously integrated thin films for either photon generation,[14,41,42] detection,[13] or nonlinear interactions[43–45] could be placed directly on top of the optical layer with photons coupling evanescently to the rest of the integrated photonics. The AlN-actuator based photonic platform is also entirely post-CMOS compatible and can thus be directly fabricated on fully formed CMOS integrated circuits[46] for direct biasing or control of hybrid systems;[47–49] thus, the architecture presented here can be scaled to both very large photonic circuit sizes and very small photonic circuit pitches without electrical control bottlenecks.

## Acknowledgements

Major funding for this work is provided by MITRE for the Quantum Moonshot Program. A.J.M. acknowledges support from the Feodor Lynen fellowship, Humboldt Foundation, and the DARPA ONISQ program. D.E. acknowledges partial support from Brookhaven National Laboratory, which is supported by the U.S.~Department of Energy, Office of Basic Energy Sciences, under Contract No.~DE-SC0012704. M.E. performed this work, in part, with funding from the Center for Integrated Nanotechnologies, an Office of Science User Facility operated for the U.S. Department of Energy Office of Science. M.D. thanks Hugo Larocque for the transmission data at 1550 nm. M.D. and M.Z. thank MITRE engineers Lilia Chan and Kevin Dauphinais for their support designing the PMMI experimental setup. M.D. also thanks Mihika Prabhu, Saumil Bandyopadhyay, Ryan Hamerly, and Carlos Errando-Herranz for helpful technical discussions.

## Author contributions

M.D. designed the experimental setups and performed the experiments, with significant assistance from G.C. for the cryogenic experiments and M.Z. for the PMMI characterization. A.J.L. and M.E. developed and supervised the entire fabrication process with assistance from D.D.; M.E. and M.D. designed the basic modulators; and M.D. laid out the design mask with assistance from G.C., A.J.L. and M.E. on the development of design and process rules. A.J.M. and D.H. assisted in photonic device characterization and design analysis. M.D., D.E., and M.E. conceived the experiment. G.G., D.E., and M.E. supervised the project. The Quantum Moonshot Program is supervised by G.G. All authors contributed to writing the paper.

## Additional information

Supplementary information is available.

## Competing interests

The authors declare no competing financial interests.

## References

1.  Bogaerts, W. *et al.* Programmable photonic circuits. *Nature* **586**, 207–216 (2020).

2.  Harris, N. C. *et al.* Linear programmable nanophotonic processors. *Optica* vol. 5 1623 (2018).

3.  Pérez, D., Gasulla, I., Mahapatra, P. D. & Capmany, J. Principles, fundamentals, and applications of

programmable integrated photonics. *Advances in Optics and Photonics* vol. 12 709 (2020).

4.  Wetzstein, G. *et al.* Inference in artificial intelligence with deep optics and photonics. *Nature* **588**, 39–47 (2020).

5.  Shen, Y., Harris, N. C., Skirlo, S., Englund, D. & Soljacic, M. Deep learning with coherent nanophotonic circuits. *2017 IEEE Photonics Society Summer Topical Meeting Series (SUM)* (2017) doi:10.1109/phosst.2017.8012714.

6.  Hamerly, R., Bernstein, L., Sludds, A., Soljačić, M. & Englund, D. Large-Scale Optical Neural Networks Based on Photoelectric Multiplication. *Physical Review X* vol. 9 (2019).

7.  Prabhu, M. *et al.* Accelerating recurrent Ising machines in photonic integrated circuits. *Optica* vol. 7 551 (2020).

8.  Levine, H. *et al.* Parallel Implementation of High-Fidelity Multiqubit Gates with Neutral Atoms. *Phys. Rev. Lett.* **123**, 170503 (2019).

9.  Kielpinski, D., Monroe, C. & Wineland, D. J. Architecture for a large-scale ion-trap quantum computer. *Nature* **417**, 709–711 (2002).

10. Choi, H., Pant, M., Guha, S. & Englund, D. Percolation-based architecture for cluster state creation using photon-mediated entanglement between atomic memories. *npj Quantum Information* vol. 5 (2019).

11. Mehta, K. K. *et al.* Integrated optical addressing of an ion qubit. *Nat. Nanotechnol.* **11**, 1066–1070 (2016).

12. Najafi, F. *et al.* On-chip detection of non-classical light by scalable integration of single-photon detectors. *Nat. Commun.* **6**, 5873 (2015).

13. Steinhauer, S. *et al.* NbTiN thin films for superconducting photon detectors on photonic and two-dimensional materials. *Applied Physics Letters* vol. 116 171101 (2020).

14. Wan, N. H. *et al.* Large-scale integration of artificial atoms in hybrid photonic circuits. *Nature* **583**, 226–231 (2020).

15. Kim, J.-H., Aghaeimeibodi, S., Carolan, J., Englund, D. & Waks, E. Hybrid integration methods for on-chip quantum photonics. *Optica* vol. 7 291 (2020).

16. Elshaari, A. W., Pernice, W., Srinivasan, K., Benson, O. & Zwiller, V. Hybrid integrated quantum photonic circuits. *Nature Photonics* vol. 14 285–298 (2020).

17. Arrazola, J. M. *et al.* Quantum circuits with many photons on a programmable nanophotonic chip. *Nature* **591**,

54–60 (2021).

18. Reck, M., Zeilinger, A., Bernstein, H. J. & Bertani, P. Experimental realization of any discrete unitary operator. *Phys. Rev. Lett.* **73**, 58–61 (1994).

19. Clements, W. R., Humphreys, P. C., Metcalf, B. J., Steven Kolthammer, W. & Walsmley, I. A. Optimal design for universal multiport interferometers. *Optica* vol. 3 1460 (2016).

20. Harris, N. C. *et al.* Quantum transport simulations in a programmable nanophotonic processor. *Nature Photonics* vol. 11 447–452 (2017).

21. Carolan, J. *et al.* Universal linear optics. *Science* vol. 349 711–716 (2015).

22. Taballione, C. *et al.* 8×8 reconfigurable quantum photonic processor based on silicon nitride waveguides. *Opt. Express* **27**, 26842–26857 (2019).

23. Baehr-Jones, T. *et al.* Ultralow drive voltage silicon traveling-wave modulator. *Opt. Express* **20**, 12014–12020 (2012).

24. Chakraborty, U. *et al.* Cryogenic operation of silicon photonic modulators based on the DC Kerr effect. *Optica* vol. 7 1385 (2020).

25. Stanfield, P. R., Leenheer, A. J., Michael, C. P., Sims, R. & Eichenfield, M. CMOS-compatible, piezo-optomechanically tunable photonics for visible wavelengths and cryogenic temperatures. *Opt. Express* **27**, 28588–28605 (2019).

26. Liu, J. *et al.* Monolithic piezoelectric control of soliton microcombs. *Nature* **583**, 385–390 (2020).

27. Tian, H. *et al.* Hybrid integrated photonics using bulk acoustic resonators. *Nat. Commun.* **11**, 3073 (2020).

28. Jin, W. *et al.* Piezoelectric tuning of a suspended silicon nitride ring resonator. *2017 IEEE Photonics Conference (IPC)* (2017) doi:10.1109/ipcon.2017.8116029.

29. Harris, N. C. *et al.* Efficient, compact and low loss thermo-optic phase shifter in silicon. *Opt. Express* **22**, 10487–10493 (2014).

30. Errando-Herranz, C. *et al.* MEMS for Photonic Integrated Circuits. *IEEE Journal of Selected Topics in Quantum Electronics* vol. 26 1–16 (2020).

31. Gyger, S. *et al.* Reconfigurable photonics with on-chip single-photon detectors. *Nat. Commun.* **12**, 1408 (2021).

32. Huang, M. Stress effects on the performance of optical waveguides. *International Journal of Solids and Structures* vol. 40 1615–1632 (2003).

33. Johnson, S. G. *et al.* Perturbation theory for Maxwell's equations with shifting material boundaries. *Physical Review E* **65**, (2002).

34. Neculoiu, D., Bunea, A.-C., Dinescu, A. M. & Farhat, L. A. Band Pass Filters Based on GaN/Si Lumped-Element SAW Resonators Operating at Frequencies Above 5 GHz. *IEEE Access* vol. 6 47587–47599 (2018).

35. Olsson, R. H. *et al.* A high electromechanical coupling coefficient SH0 Lamb wave lithium niobate micromechanical resonator and a method for fabrication. *Sensors and Actuators A: Physical* vol. 209 183–190 (2014).

36. Atatüre, M., Englund, D., Vamivakas, N., Lee, S.-Y. & Wrachtrup, J. Material platforms for spin-based photonic quantum technologies. *Nature Reviews Materials* vol. 3 38–51 (2018).

37. Teshigahara, A., Hashimoto, K.-Y. & Akiyama, M. Scandium aluminum nitride: Highly piezoelectric thin film for RF SAW devices in multi GHz range. *2012 IEEE International Ultrasonics Symposium* (2012) doi:10.1109/ultsym.2012.0481.

38. West, G. N. *et al.* Low-loss integrated photonics for the blue and ultraviolet regime. *APL Photonics* vol. 4 026101 (2019).

39. Lu, T.-J. *et al.* Aluminum nitride integrated photonics platform for the ultraviolet to visible spectrum. *Opt. Express* **26**, 11147–11160 (2018).

40. Fan, L., Sun, X., Xiong, C., Schuck, C. & Tang, H. X. Aluminum nitride piezo-acousto-photonic crystal nanocavity with high quality factors. *Applied Physics Letters* vol. 102 153507 (2013).

41. Ellis, D. J. P. *et al.* Independent indistinguishable quantum light sources on a reconfigurable photonic integrated circuit. *Applied Physics Letters* vol. 112 211104 (2018).

42. Papon, C. *et al.* Nanomechanical single-photon routing. *Optica* vol. 6 524 (2019).

43. Zhu, D. *et al.* Integrated photonics on thin-film lithium niobate. *Advances in Optics and Photonics* vol. 13 242 (2021).

44. Desiatov, B., Shams-Ansari, A., Zhang, M., Wang, C. & Lončar, M. Ultra-low-loss integrated visible photonics using thin-film lithium niobate. *Optica* vol. 6 380 (2019).

45. Cai, L. *et al.* Acousto-optical modulation of thin film lithium niobate waveguide devices. *Photonics Research* vol. 7 1003 (2019).

46. Wojciechowski, K. E., Olsson, R. H., Tuck, M. R., Roherty-Osmun, E. & Hill, T. A. Single-chip precision oscillators based on multi-frequency, high-Q aluminum nitride MEMS resonators. *TRANSDUCERS 2009 - 2009 International Solid-State Sensors, Actuators and Microsystems Conference* (2009) doi:10.1109/sensor.2009.5285626.

47. Patra, B. *et al.* Cryo-CMOS Circuits and Systems for Quantum Computing Applications. *IEEE Journal of Solid-State Circuits* vol. 53 309–321 (2018).

48. Kim, D. *et al.* A CMOS-integrated quantum sensor based on nitrogen–vacancy centres. *Nature Electronics* vol. 2 284–289 (2019).

49. Ibrahim, M. I., Foy, C., Englund, D. R. & Han, R. High-Scalability CMOS Quantum Magnetometer With Spin-State Excitation and Detection of Diamond Color Centers. *IEEE Journal of Solid-State Circuits* vol. 56 1001–1014 (2021).

**Methods**

**Holding power consumption and reconfiguration energy calculations.** We first calculate the device capacitance and on-chip routing metal resistance based upon the measured RC roll-off in our frequency response curve, finding $C = 17$ pF and $R_{chip} = 30$ Ω in addition to a series voltage-source resistance of 50 Ω. The device leakage resistance based on previous measurements[25] is estimated to be 500 MΩ and 200 GΩ at 300 K and 5 K respectively. The holding power $P = IV$ is found for two phase-shifters holding 25 V to maintain a π phase-shift. The reconfiguration energy dissipated on-chip is calculated via the formula $E = (R_{chip}/R_{tot})CV^2$ where $R_{tot}$ is the total series resistance of the circuit including external resistors in the AWG and $V_s = 50$ V. This equation accounts for two phase-shifters with a 50 V swing. To reduce on-chip energy dissipation for slower reconfiguration rates, $R_{tot}$ is adjusted such that the corner frequency $1/(2\pi R_{tot} C)$ matches the current reconfiguration frequency (which is done off-chip at the voltage source). This minimizes the on-chip energy dissipation while maintaining the necessary reconfiguration rate. The total power dissipated (plotted in Fig. 1b) is simply the holding power plus the reconfiguration energy times the reconfiguration rate.

**Device characterization.** We characterize individual MZI modulators both at room and cryogenic temperatures with 250-um pitch optical fiber arrays grating coupled to the on-chip waveguides. We use a 150-um pitch RF probe (in GSGSG configuration for room temperature, GSG for cryogenic temperature) to apply high-frequency electrical signals.

**Digital lock-in amplifier.** A time-trace of both output channels 5 and 6 of the PMMI is digitized directly by a high-speed oscilloscope. The time-traces are digitally integrated (for 4 ms) with sinusoids of varying frequencies and phases, whose resulting amplitudes form the data for the plots in Fig. 5e, f. The two photodiodes used for both channels had 125 MHz and 600 MHz bandwidths respectively.

**Data availability.** The data that support the plots within this paper are available from the corresponding authors upon reasonable request.

Please see Supplementary Information for more detailed experimental methods.

# High-speed programmable photonic circuits in a cryogenically compatible, visible-NIR 200 mm CMOS architecture: Supplementary Information

Mark Dong,[1,2,6] Genevieve Clark,[1,2] Andrew J. Leenheer,[3] Matthew Zimmermann,[1] Daniel Dominguez,[3] Adrian J. Menssen,[2] David Heim,[1] Gerald Gilbert,[2,4,7] Dirk Englund,[2,5,8] and Matt Eichenfield[3,9]

[1]The MITRE Corporation, 202 Burlington Road, Bedford, Massachusetts 01730, USA
[2]Research Laboratory of Electronics, Massachusetts Institute of Technology, Cambridge, Massachusetts 02139, USA
[3]Sandia National Laboratories, P.O. Box 5800 Albuquerque, New Mexico, 87185, USA
[4]The MITRE Corporation, 200 Forrestal Road, Princeton, New Jersey 08540, USA
[5]Brookhaven National Laboratory, 98 Rochester St, Upton, New York 11973, USA
[6]mdong@mitre.org
[7]ggilbert@mitre.org
[8]englund@mit.edu
[9]meichen@sandia.gov

## 1. Wafer Fabrication

**Table S1: Nominal film thicknesses**

| Process Material | M1 (Al) | a-Si | W | M2 (Al) | AlN | M3 (Al) | SiN$_x$ |
|---|---|---|---|---|---|---|---|
| Nominal thickness (nm) | 700 | 200 | 560 | 100 | 450 | 250 | 300 |

We fabricated all devices at Sandia National Laboratories. The nominal film thicknesses with labels corresponding to Fig. 2c are listed in Table S1. Starting with an oxide film on silicon, we sequentially deposited and patterned the M1 routing (Al/Ti), amorphous silicon (a-Si) release layer, tungsten vias, M2 bottom electrode, AlN, M3 top electrode, bottom oxide cladding, SiN, and top oxide cladding. Several chemical-mechanical polishing steps were introduced to ensure wafer flatness and more uniform film thicknesses, especially before critical layers such as M2 for AlN growth and SiN for reliable waveguide etching. We then performed an oxide etch to open all metal pads for probe or wire-bond access. To complete the fabrication, we etched deep trenches through all layers until reaching the a-Si to define all piezo-optomechanical structures, then subsequently performed a XeF$_2$ release, etching away the a-Si and undercutting all devices.

We note several major improvements in our fabrication since our earlier work:[25] all processes have been recalibrated to operate on 200 mm wafer technology (from previous 150 mm); we greatly improved the deep trench process and utilized narrower trenches for flatter device topography; and we added additional CMP processes for more uniform definitions of pillars and trenches.

## 2. Optical loss characterization

We characterize the propagation loss in our single-mode SiN waveguides with simple ring-resonators by extracting quality factors at both visible (780 nm) and NIR (1550 nm) wavelengths. As seen in Fig. S1, we estimate the intrinsic Q to be ~1.6 million at 780 nm and ~30000 at 1550 nm, corresponding to propagation losses of -0.39 dB/cm and -9.1 dB/cm respectively.

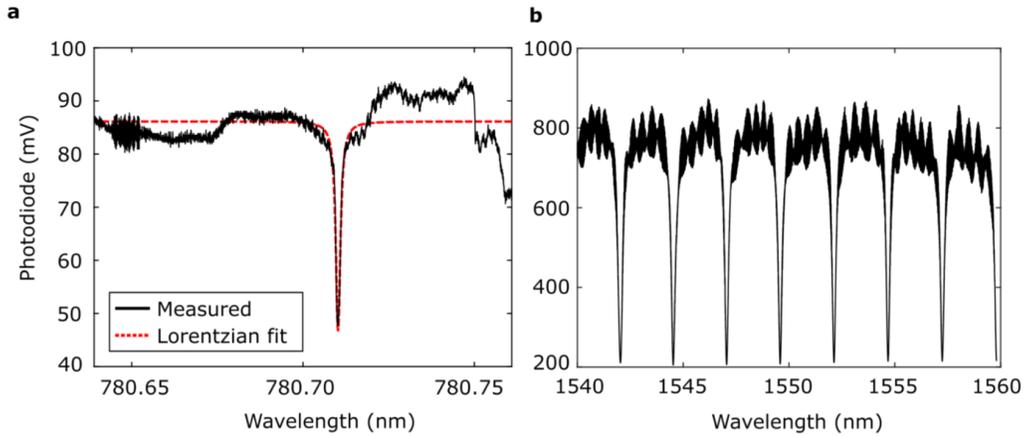

**Fig. S1: Ring resonances at 780 nm and 1550 nm.** a) scan of a single resonance near 780 nm with Lorentzian fit b) wide scan of several resonances at 1550 nm

Using individual loop-back test structures with varying numbers of Euler curve loops, we measure the loss per loop for a 4-um minimum radius, 180 degree Euler curve to be -0.21 dB $\pm$ 0.042 per bend. The single-mode to multimode waveguide adiabatic tapers are also separately characterized in loop-back structures with varying number of tapers, resulting in a loss of -0.022 dB $\pm$ 0.038 dB per taper. Fig. S2a shows a subset of our loss measurements in the bends and Fig. S2b shows the total coupled MZI power when actuating, with very little power modulation when tuning the phase-shifters. The standard deviation of the total coupled power is roughly 1.1% of the total coupled power.

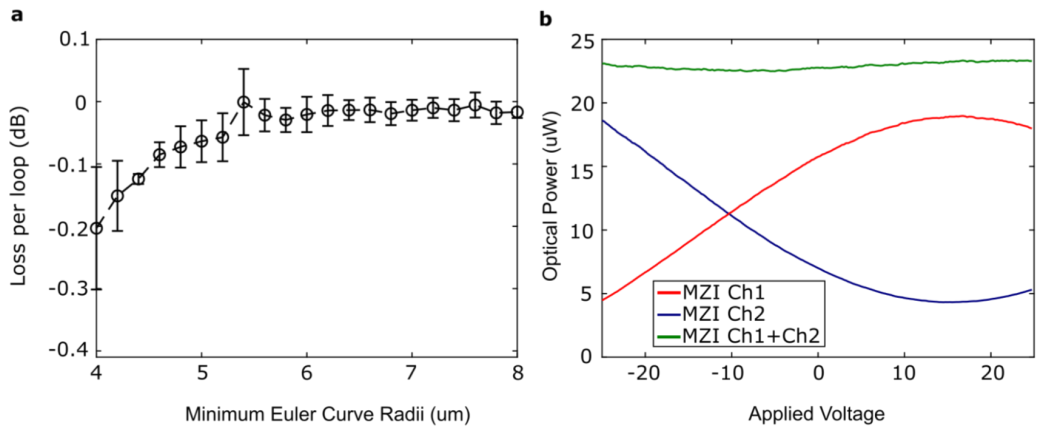

**Fig. S2: Bending and actuation losses.** a) loss per euler curve loop in dB as the minimum bending radii is increased b) optical power plots at 737 nm of the MZI modulator showing the total power during modulation is roughly constant

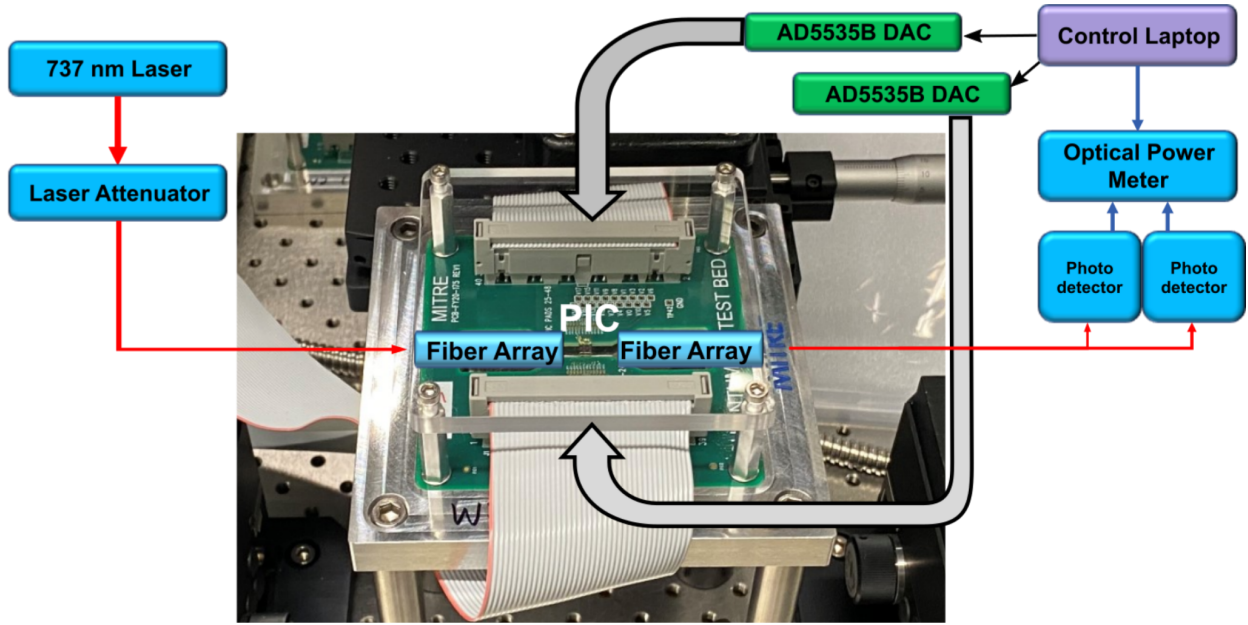

**Fig S3: Diagram of the PNP experimental setup.** The interface between the PIC, control electronics, and optical components is displayed.

Propagation loss in the 5 um waveguide actuation region is measured using varying length phase-shifters with the same number of waveguide bends and adiabatic tapers. Measurements of the 5 um propagation loss did not have a large sample size and fitted losses are found to be -0.116 $\pm$ 0.75 dB/cm.

### 3.     PMMI test setup hardware

An overview of our PMMI experimental setup is shown in Fig. S3. We send 737 nm laser power into one channel of the input fiber array per sweep, while monitoring two outputs at a time. Cutouts in our custom-built PCB and chassis are used to allow custom fiber array holders to edge couple an 8-channel fiber array to our 4x4 PMMI, as shown in Fig. S4. The edge coupling is currently unoptimized, with efficiencies of - 15 dB per facet.

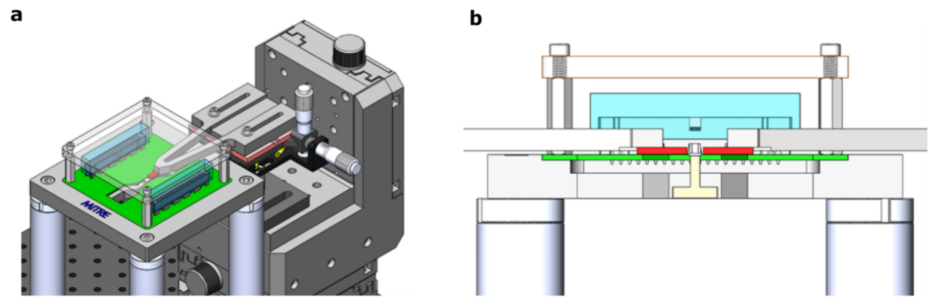

**Fig. S4: CAD renders of our mechanical components for the PNP setup.** a) isometric view of a mechanical model of the chip testing setup b) side view of chip testing setupt

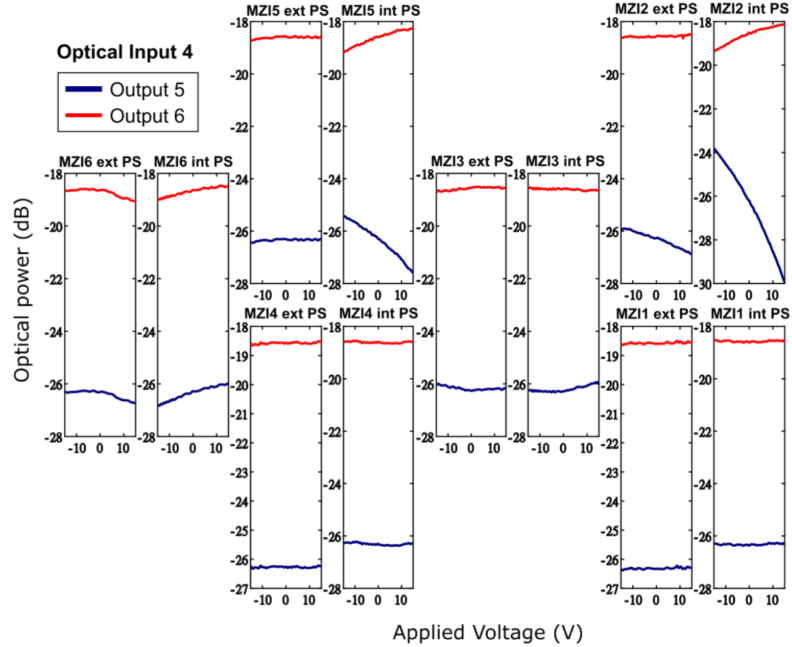

**Fig. S5: Dataset of all a baseline voltage sweep of all twelve phase-shifters of our 4x4 PNP.** MZI labels follow those of Fig. 5. The optical input for this data is 4 while monitoring outputs 5,6.

Electronic control of the PICs is done using two Analog Devices 5535B Evaluation Boards. These boards use the AD5535B 32-Channel, 14-bit DAC with on-chip high voltage output amplifiers which are capable of amplifying each channel up to 200 V at up to 550 µA of driving current. The evaluation kit routes all channels to an IDC40 40-pin connector, so an IDC40 ribbon cable connects the evaluation boards to a custom chip carrier PCB. The two AD5535B Evaluation Boards are each connected to the Analog Devices SDP-S System Demonstration Platform board, which allows voltages to be set on the AD5535B DAC over Serial Peripheral Interface (SPI) and is connected to a control laptop via a mini-USB cable. Power is supplied to the board from a 5 V supply for digital logic and low-voltage analog power and a 200 V supply for the amplifier. When set to a maximum output voltage of 200V, the AD5535B will have a minimum step size of 12.5 mV. The channels can exhibit crosstalk of up to 4 volts, so a delay of 250 ms is used after setting a voltage to allow signals and crosstalk effects time to settle before making a measurement of the optical response.

Custom python drivers are written to set voltages on each channel from a control laptop, and an additional layer of software is written to allow a user to set voltages with respect to the individual devices on the 4x4 PMMI. We run top-level scripts utilizing these drivers to sweep the actuation voltage of devices while measuring the response on an optical power meter.

A custom printed circuit board and chassis are designed to hold the 4x4 PMMI and secure it for edge coupling on an optics table. The PCB routed control voltages and ground signals from a 2x20 pin connector to wirebond pads, which were then wirebonded to the PIC. The PIC was placed on a copper mount in the middle of the board.

### 4. PMMI calibration

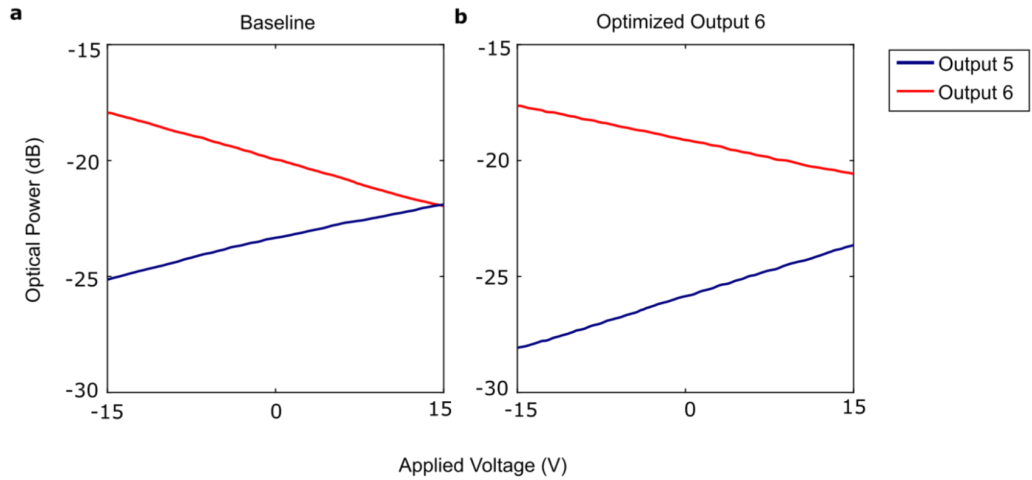

**Fig. S6: Sample calibration plots.** a) baseline and b) channel 6 optimized results when sweeping MZI5

All software used to sweep the voltage is programmed in python and automated scripts are run to set each phase shifter. Because our control electronics only allow one voltage polarity, we sweep phase-shifters in each arm in an opposite manner. When sweeping an MZI modulator, the first arm is swept from a positive voltage down to 0V (hence the negative voltage in the plot), then the second arm is swept from 0V to a positive voltage. When sweeping, voltages are set in 0.25 V steps and the optical power level from two fiber outputs are recorded on a benchtop power meter.

Initially, phase-shifters have their voltages swept while keeping the voltage on all other phase-shifters at 0V. These are recorded as 'baseline' power measurements (see Fig. S5 for an example baseline sweep of all twelve phase-shifters on the PMMI). After baseline measurements are made, the voltages on all devices are set to values

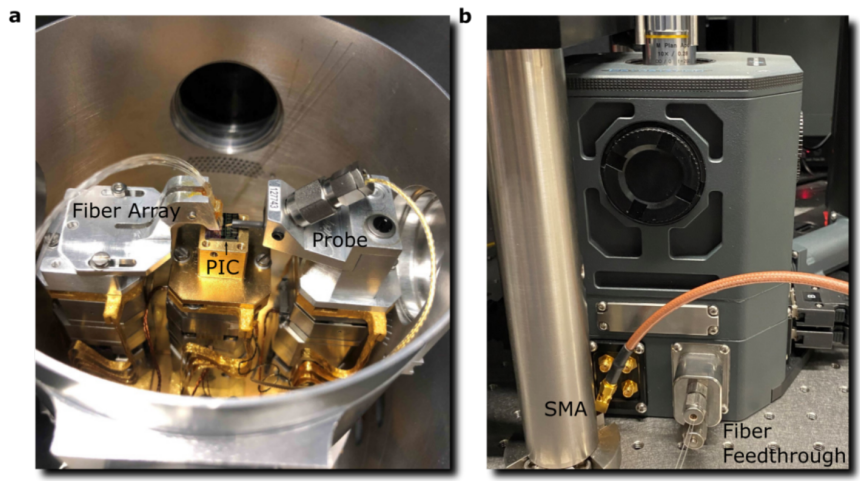

**Fig. S7: Pictures of the cryostat setup.** a) inside the chamber showing the fiber array, PIC, and RF probe each on its own nanopositioner b) outside the chamber showing the SMA and fiber feedthroughs

that would maximize the power output of one measured channel before sweeping the voltage on one specific phase shifter. These values are recorded as the "optimized" power measurements, with a comparison shown in Fig. S6. Characterization sweeps are limited from -15V to +15V to avoid the breakdown of the vias at high voltages. We note that when changing the input or output channels, the optical test setup is realigned to maximize the measured power before collecting new baseline and optimized measurements.

## 5. Cryogenic test setup

We perform cryogenic testing of our photonics in a Montana, closed-cycle cryostat equipped with a three nanopositioner stack. The PIC is cooled down to ~5.1 K and optical and electrical I/O are provided with a fiber array and RF probe (Fig. S7a). The optical and electrical connections are routed through the chamber via built-in feedthroughs ((Fig. S7b). The monitored temperature remained at roughly 5.1 K throughout the entire duration of the cryogenic characterization measurements, including DC and AC actuation tests.